\documentclass[prd,preprint,tightenlines,superscriptaddress,floatfix,showpacs,preprintnumbers,nofootinbib,eqsecnum]{revtex4}

 \usepackage[dvips,final]{graphicx}
     \usepackage{epsfig}
      \usepackage{bm}

\newcommand{\bq}{\mbox{\boldmath $q$}}

\newcommand{\bs}{\mbox{\boldmath $s$}}

\newcommand{\bb}{\mbox{\boldmath $b$}}

\newcommand{\bE}{\mbox{\boldmath $E$}}
\def\Pom{I\!\!P}
\def\Reg{I\!\!R}

\def\lsim{\mathrel{\rlap{\lower4pt\hbox{\hskip1pt$\sim$}}
    \raise1pt\hbox{$<$}}}         
\def\gsim{\mathrel{\rlap{\lower4pt\hbox{\hskip1pt$\sim$}}
    \raise1pt\hbox{$>$}}}

\begin{document}

\thispagestyle{empty} \preprint{\hbox{}} \vspace*{-10mm}

\title{Exclusive production of
$\rho^0 \rho^0$ pairs in $\gamma \gamma$ collisions
at RHIC}

\author{M.~K\l usek}
\email{mariola.klusek@ifj.edu.pl}

\affiliation{University of Rzesz\'ow, PL-35-959 Rzesz\'ow,
Poland}
\affiliation{Institute of Nuclear Physics PAN, PL-31-342 Cracow, Poland}

\author{W.~Sch\"afer}
\email{wolfgang.schafer@ifj.edu.pl}
\affiliation{Institute of Nuclear Physics PAN, PL-31-342 Cracow,
Poland} 

\author{A.~Szczurek}
\email{antoni.szczurek@ifj.edu.pl}
\affiliation{University of Rzesz\'ow, PL-35-959 Rzesz\'ow,
Poland}
\affiliation{Institute of Nuclear Physics PAN, PL-31-342 Cracow,
Poland}

\date{\today}

\begin{abstract}
We discuss exclusive electromagnetic production of 
two neutral $\rho$ mesons in coherent photon-photon
processes in ultrarelativistic heavy-ion collisions.
The cross section is calculated in the equivalent photon
approximation (EPA).
Both uncertainties of the flux factors and photon-photon
cross sections are discussed in details.
We show that inclusion of precise charge densities
in nuclei is essential for realistic evaluations of
the nuclear photon-photon cross sections.
We find that the cross section, especially with realistic
flux factors, is sensitive to low
energy in the subsystem $\gamma \gamma \to \rho^0 \rho^0$.
The experimental data for the
$\gamma \gamma \to \rho^0 \rho^0$ cross section 
extracted from $e^+ e^-$ collisions are parametrized and 
used to estimate the nucleus-nucleus cross section.
In addition, we include vector-dominance-model(VDM)--Regge contribution 
which becomes important at large photon-photon energy.
Large nuclear cross sections are obtained.
We discuss a possibility of focusing on the large-energy
component. 
We find that both $\rho^0$ mesons are 
produced predominantly at midrapidities and could 
be measured by the STAR collaboration at RHIC. 
\end{abstract}


\maketitle

\section{Introduction}

Exclusive production of elementary particles (lepton pairs,
Higgs, etc.) or mesons (vector mesons, pair of pseudoscalar
mesons, etc.) in ultrarelativistic heavy ion collisions is 
an interesting 
and quickly growing field \cite{BGMS75,BHTSK02,Hencken} of 
theoretical investigation.
On experimental side the situation is slightly different.
So far only single-$\rho^0$ exclusive cross section
$A A \to A A \rho^0$ was measured \cite{STAR_rho0}. 
Here the dominant mechanism is the photoproduction
described by the photon-pomeron fusion.

\begin{figure}[!h]    
\includegraphics[width=0.35\textwidth]{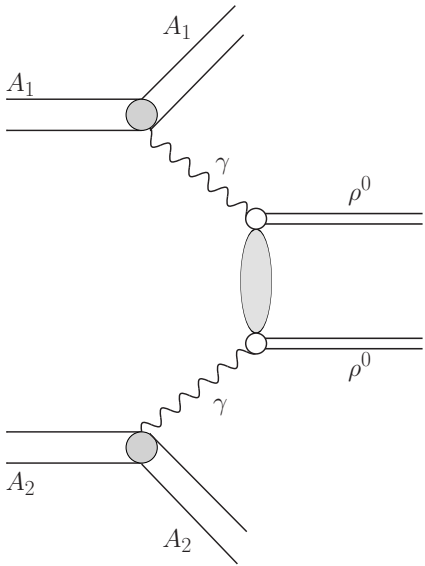}
   \caption{\label{fig:diagram}
   \small 
The reaction discussed in this paper.
}
\end{figure}

In the present paper we consider exclusive production 
of $\rho^0 \rho^0$ pairs in $\gamma \gamma$ collisions.
So far only integrated cross section for 
$A A \to A A \rho^0 \rho^0$ was roughly estimated in 
the literature \cite{Goncalves}, using a pQCD--inspired modelling
of the Pomeron exchange between mesons for 
$A A \to A A J/\Psi J/\Psi$ and 
$A A \to A A \rho^0 J/\Psi$ and
assuming Regge factorization.
Uncertainties of almost three-orders of magnitude
were found.
As we discuss here, in the $A A \to A A \rho^0 \rho^0$
reaction one is sensitive rather 
to the close-to-threshold region for the 
$\gamma \gamma \to \rho^0 \rho^0$ subprocess where the 
specific pQCD approaches of \cite{Goncalves}
are not applicable at all.
Furthermore the Regge-factorization for this hard and soft
processes is likely broken.
The $\rho^0 \rho^0$ pair will eventually decay into the final state 
of four charged pions. If the $\rho^0 \rho^0$ invariant mass is
very large, the two $\pi^+ \pi^-$--pairs will be separated 
by a large rapidity gap. It has been recently suggested that in this 
kinematics, charge asymmetries can
be used to discover/study the QCD-Odderon 
exchange \cite{Pire}.

At the RHIC energies discussed here, we will be however dominated 
by  $\gamma \gamma$ center-of-mass energies in the region of a few GeV.
Here, experimental data for
$\gamma \gamma \to \rho^0 \rho^0$ are known and were
measured by several groups at $e^+ e^-$ colliders
\cite{original_data}. Experimentally one 
observes a huge enhancement close to the threshold.
The origin of this enhancement was never understood.
Some speculations are discussed in Ref.\cite{Rosner2004}.
The huge cross section at $W \approx$ 1.5-2 GeV
was interpreted as a tensor resonance decaying into
$\rho^0 \rho^0$ channel \cite{Anikin}.
In Ref.\cite{Levy} the Regge factorization was tested.
In the following we shall see the consequences
of the anomalous behaviour seen in the $e^+ e^-$ data 
for the two-photon production of the $\rho^0 \rho^0$ pair
in the nucleus-nucleus collisions.

\section{Calculation}

\subsection{Equivalent photon approximation for nuclei}

Let us consider the process $A A \to A A \rho^0 \rho^0$
depicted in Fig.\ref{fig:diagram}.
The cross section takes the familiar form of a convolution of
equivalent photon fluxes and $\gamma \gamma$--cross sections:

\begin{eqnarray}
{d\sigma(AA \to \rho^0\rho^0 AA; s_{AA}) \over d^2 \bb} = 
dn_{\gamma \gamma}(x_1,x_2,\bb) \, {\hat \sigma}(\gamma \gamma \to \rho^0 \rho^0; 
x_1 x_2 s_{AA}) + \dots
\label{AA_xsec}
\end{eqnarray}
Here we omitted helicity--dependent pieces, see below for more details.
The effective photon flux is expressed through the electric field strengths
of the ions and reads \cite{Baur_absorption} (see also \cite{CJ90}):

\begin{eqnarray}
dn_{\gamma \gamma} (x_1,x_2,\bb) = 
\int d^2\bb_1 d^2\bb_2  \, S^2_{abs}(\bb) \delta^{(2)}(\bb - \bb_1 + \bb_2)
{dx_1 \over x_1} {dx_2 \over x_2} {1 \over \pi^2} |\bE(x_1,\bb_1)|^2 
|\bE(x_2,\bb_2)|^2 \, , \nonumber \\
\label{flux}
\end{eqnarray}
where the electromagnetic field strengths are given in terms of the 
charge form factor of the nucleus $F_{em}(\vec{q}^2)$ by:

\begin{eqnarray}
\bE(x,\bb) &&= Z \sqrt{4 \pi \alpha_{em}}
\int {d^2\bq \over (2 \pi )^2} \exp[-i \bb \bq] \,  \, {\bq \over \bq^2 + x^2M_A^2} \, 
F_{em}(\bq^2 + x^2 M_A^2 ) \, \nonumber \\
&&= -i  Z \sqrt{4 \pi \alpha_{em}}
\, {\bb \over b^2} \, \int {d \bq^2 \over 4 \pi} {(b q)J_1(b q) 
\over \bq^2 + x^2 M_A^2} F_{em}(\bq^2 + x^2 M_A^2 ) \, .
\label{E-field}
\end{eqnarray}

In practice $F_{em}(\vec{q}^2)$ is obtained as the
Fourier transform of the nuclear charge densities extracted
from electron--nucleus scattering data, for a useful compilation, see 
\cite{Barrett_Jackson,VJV87}.
By putting $F_{em} \equiv 1$ one obtains the well-known result for 
a point-like charge (see e.g. the textbook \cite{Jackson}):

\begin{eqnarray}
\bE_{pt}(x,\bb) = -i \, {Z \sqrt{4 \pi \alpha_{em}}  \over 2 \pi} 
\, {\bb \over b^2} \, (xM_A b) K_1(xM_A b)
\, .
\label{pointlike}
\end{eqnarray}

Notice that our formula for the effective Weizs\"acker --Williams flux of photons includes the impact-parameter 
dependent absorption factor $S_{abs}^2(\bb)$.
It represents the probability that no inelastic interaction between
the nuclei occurs. At RHIC energies it could be for example estimated
from Glauber theory using the Czy\.z--Maximon \cite{Czyz} approximation:

\begin{eqnarray}
S_{abs}^2(\bb) = \exp\Big( - \sigma^{tot}_{NN} \, \int d^2 \bs 
\, T_A (\bb - \bs) T_A(\bs) \Big) \, ,
\end{eqnarray}
where $T_A(\bb)$ is the optical thickness of the nucleus.
We checked, that for our purposes it can be well approximated by the result 
for a black disc:
\begin{equation}
S^2_{abs}(\bb) = \theta(b - 2R_A) \, .
\end{equation}

The presence of the absorption factor also induces an azimuthal correlation
between impact parameters $\bb_1$ and $\bb_2$. Now, soft photons are polarized
linearly in the transverse plane, and therefore a subtle helicity dependence
arises. Strictly speaking, one should write
\begin{eqnarray}
{d\sigma(AA \to \rho^0\rho^0 AA; s_{AA}) \over d^2 \bb} = 
dn^{\parallel}_{\gamma \gamma}(x_1,x_2,\bb) \, 
{\hat \sigma}_\parallel(\gamma \gamma \to \rho^0 \rho^0) +
 dn^{\perp}_{\gamma \gamma}(x_1,x_2,\bb) \, 
{\hat \sigma}_\perp(\gamma \gamma \to \rho^0 \rho^0) \, ,
\nonumber \\
\end{eqnarray}
where 
\begin{eqnarray}
dn^\parallel_{\gamma \gamma} (x_1,x_2,\bb) &=& 
\int d^2\bb_1 d^2\bb_2  \, S^2_{abs}(\bb) \delta^{(2)}(\bb - \bb_1 + \bb_2)
{dx_1 \over x_1} {dx_2 \over x_2} {1 \over \pi^2} |\bE(x_1,\bb_1) \cdot 
\bE(x_2,\bb_2)|^2 \, , \nonumber \\
dn^\perp_{\gamma \gamma} (x_1,x_2,\bb) &=& 
\int d^2\bb_1 d^2\bb_2  \, S^2_{abs}(\bb) \delta^{(2)}(\bb - \bb_1 + \bb_2)
{dx_1 \over x_1} {dx_2 \over x_2} {1 \over \pi^2} |[\bE(x_1,\bb_1) \times
\bE(x_2,\bb_2)]|^2 \, , \nonumber \\
\end{eqnarray}
are the fluxes for photons with parallel ($\parallel$), and 
perpendicular ($\perp$) linear polarizations; 
$\sigma^\parallel$ and $\sigma^\perp$
are the corresponding $\gamma \gamma$ cross sections.
The small effects associated with this helicity dependence have been 
discussed in \cite{Baur_absorption} and will be neglected here.

A further comment on the practical uses of the equivalent 
photon approximation is in order.
It is often pointed out that the Fermi--Weizs\"acker--Williams
idea of the equivalent photon flux represents indeed the archetypical
parton--model concept. 
In this spirit, often flux factors of equivalent photons
are calculated as for point-like particles with charge $Z e$,
and the total cross section is then evaluated using
a simple parton--model type formula:
\begin{equation}
\sigma \left( AA \to A (\rho^0 \rho^0) A \right) =
\int d \omega_1 d \omega_2 
\frac{n(\omega_1)}{\omega_1}
\frac{n(\omega_2)}{\omega_2}
\hat{\sigma} \left( \gamma \gamma \to \rho^0 \rho^0 \right) \; .
\label{EPA_formula}
\end{equation}
When calculating the two-dimensional integral it must 
be checked if $W^2 = 4 \omega_1 \omega_2 >  4 m_{\rho}^2$.
Here 
\begin{eqnarray}
n(\omega) \equiv \int d^2\bb \, N(\omega,\bb) = {1 \over \pi} \int d^2\bb \Big| 
\bE(x, \bb) \Big|^2 \, \;\; ; \;\; x = {\omega \over \gamma M_A} \, ,
\label{single_flux}
\end{eqnarray}
is the flux of photons in an ultrarelativistic charge of 
energy $\gamma M_A$. For RHIC, at the cms-energy 
$\sqrt{s_{NN}} = 200 \, \mathrm{GeV}$ we have $\gamma \sim 100$. 

Notice that such a factorized form is not borne out by the general
formula (\ref{flux}) which accounts for the strong absorptive corrections.
In particular, one cannot meaningfully improve the $b$--integrated
Eq.(\ref{EPA_formula}) to effectively account for strong absorption,
in particular this is \emph{not} achieved by simply restricting 
the impact parameter integrals (\ref{single_flux}) of 
the individual photon fluxes.

\begin{figure}[!h]    
\includegraphics[width=0.40\textwidth]{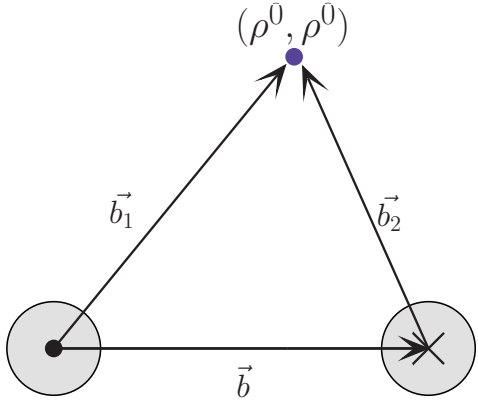}
   \caption{\label{fig:collision_in_bspace}
   \small 
The quantities used in the impact parameter calculation.
}
\end{figure}

When written in terms of photon energies $\omega_i$ in the nucleus-nucleus
center-of-mass, formula (\ref{AA_xsec}) takes the well-known form 
\cite{Baur_absorption}

\begin{eqnarray}
\sigma \left( AA \to A (\rho^0 \rho^0) A; s_{AA} \right) &=&
\int d^2 \bb_1 d^2\bb_2 
\theta \left(|\bb_1 - \bb_2| - 2R_A \right) \;
{d \omega_1 \over \omega_1} 
{d \omega_2 \over \omega_2}
N(\omega_1, \bb_1)
N(\omega_2, \bb_2)
\nonumber \\
&\times&
{\hat \sigma} 
\left( \gamma \gamma \to \rho^0 \rho^0; 4 \omega_1 \omega_2 \right) \; .
\label{bspace_EPA_formula}
\end{eqnarray}
We also use the $\gamma \gamma$ cms-energy $W_{\gamma \gamma}$, and 
the rapidity-type variable $Y$ defined through

\begin{eqnarray}
\omega_1 = {W_{\gamma \gamma}\over 2} e^Y \, , \, 
\omega_2 = {W_{\gamma \gamma} \over 2} e^{-Y} \, \,  , \, \, 
{d\omega_1 \over \omega_1} \, {d \omega_2 \over \omega_2} = 
2 \, {d W_{\gamma \gamma} \over W_{\gamma \gamma}} \, dY.
\end{eqnarray}
  
The $\sigma \left( \gamma \gamma \to \rho^0 \rho^0 \right)$
cross section can be calculated from models and/or taken 
from experimental data, if existing.
Below we discuss separately the low-$W_{\gamma \gamma}$ 
and high-$W_{\gamma \gamma}$ case.

\subsection{Low-energy $\gamma \gamma \to \rho^0 \rho^0$ cross section}

The cross section for this process was measured
up to $W_{\gamma \gamma} = 4 \, \mathrm{GeV}$ \cite{MPW}.
At low energy one observes a huge increase of the cross section.
Several possible scenarios were discussed in this context
(see e.g. Ref. \cite{Rosner2004}). 
For one real and one virtual photon induced processes
the enhancement was interpreted as due to an isotensor 
meson decay \cite{APSTW06}.
We leave the difficult problem of the microscopic origin
of the close-to-threshold bump for future studies and 
take here a pragmatic attitude of using directly
experimental data. 

\begin{figure}[!h]    
\includegraphics[width=0.4\textwidth]{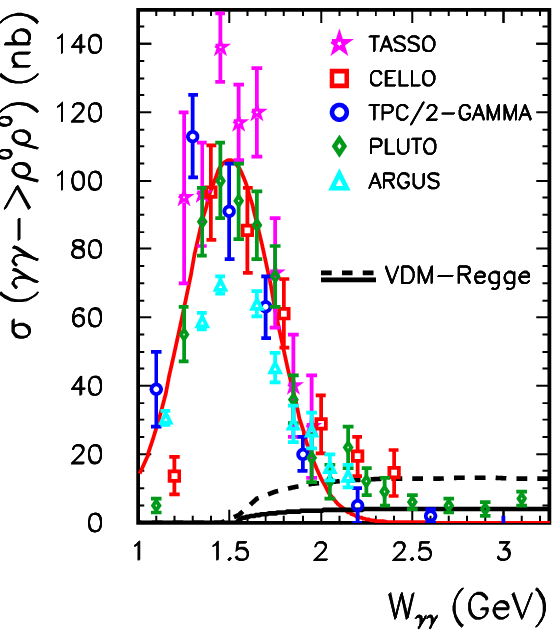}
\includegraphics[width=0.4\textwidth]{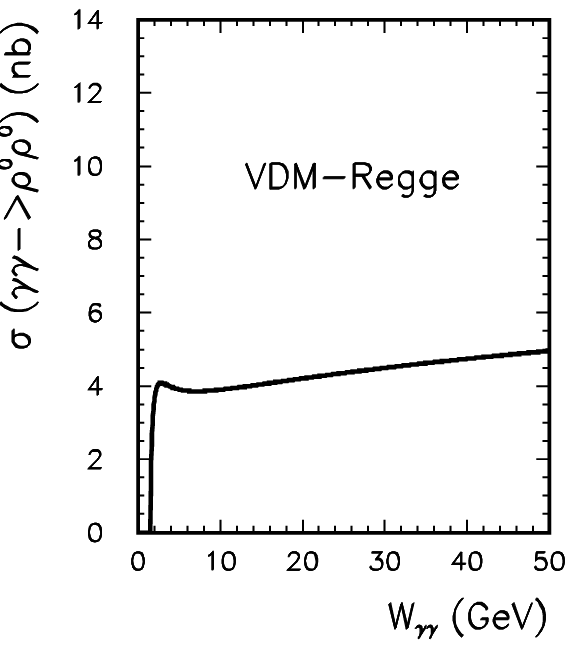}
   \caption{\label{fig:fit_gamgam_rhorho}
   \small
The elementary cross section for 
the $\gamma \gamma \to \rho^0 \rho^0$ reaction.
In the left panel we display the collection of the $e^+ e^-$ 
experimental data \cite{MPW,original_data} and our fit.
In the right panel we show our predictions based on the VDM-Regge
model decribed in the text. For comparison we show by the dashed line 
in the left panel also the result when the form factor 
correcting for off-shell effect is ignored.
}
\end{figure}

In Fig.\ref{fig:fit_gamgam_rhorho} we have collected 
the world data (see \cite{MPW,original_data} 
and references therein).
A huge rise of the cross section can be seen
close to the threshold. 
Here we shall use rather
directly experimental data in order to evaluate
the cross section in nucleus-nucleus collisions.
In Fig.\ref{fig:fit_gamgam_rhorho} we show our fit
to the world data. One can observe a small inconsistency
of the data measured by different groups.
This means that also our parametrization has about
20 \% accuracy. 

\subsection{High-energy 
$\gamma \gamma \to \rho^0 \rho^0$ cross section}

The cross section above $W_{\gamma \gamma} = 4 \, \mathrm{GeV}$ 
was never measured in the past. 
It is well known that the cross section 
for $\gamma \gamma \to \mathrm{hadrons}$ can be well described
in the VDM-Regge type model.
Here we present a similar approach but for the
specific, well defined, final state channel $\rho^0 \rho^0$.

In the VDM-Regge approach the amplitude
for the $\gamma \gamma \to \rho^0 \rho^0$ can
be written as
\begin{equation}
{\cal M}_{\gamma \gamma \to \rho^0 \rho^0}(\hat{s},\hat{t};q_1,q_2) = 
C_{\gamma \to \rho^0} C_{\gamma \to \rho^0} 
{\cal M}_{\rho^{0*} \rho^{0*} \to \rho^0 \rho^0} (\hat{s},\hat{t};q_1,q_2) \; .
\label{Regge_amplitude}
\end{equation}
Above $C_{\gamma \to \rho^0} = \sqrt{f_{\rho}^2}$ and
we use $f_{\rho}^2 = \frac{\alpha_{em}^2}{2.54}$.
The later value is obtained in order to reproduce 
the decay width of the $\rho^0$ meson into dileptons.

For energies $W_{\gamma \gamma} >$ 2-3 GeV the amplitude 
for the $\rho^{0*} \rho^{0*} \to \rho^0 \rho^0$
proceses can be written in the Regge form 
\footnote{We assume helicity conservation and neglect helicity flip 
in the present analysis. As far as only a good fit of the total
cross section is concerned this is of no further relevance for our 
purposes.}:
\begin{equation}
{\cal M}_{ \rho^{0*} \rho^{0*} \to \rho^0 \rho^0 } 
(\hat{s}, \hat{t})= \hat{s}
\left( 
\eta_{\Pom}({\hat s},{\hat t}) \; C_{\Pom} \left( \frac{\hat{s}}{s_0} \right)^{\alpha_{\Pom}({\hat t})-1} +
\eta_{\Reg}({\hat s},{\hat t}) \; C_{\Reg} \left( \frac{\hat{s}}{s_0} \right)^{\alpha_{\Reg}({\hat t})-1}
\right) \cdot F({\hat t};q_1^2) \; F({\hat t};q_2^2) \; .
\end{equation}
Above we have introduced vertex form factors
which, in general, are functions of exchanged 
pomeron/reggeon four-momentum and photon ($\rho^0$ meson)
virtualities.
We parametrize them in the factorized form:
\begin{equation}
F({\hat t};q^2) = \exp \left( \frac{B \hat{t}}{4} \right) \cdot 
\exp \left( \frac{q^2-m_{\rho}^2}{2 \Lambda^2} \right) \; .
\label{vertex_formfactor}
\end{equation}
The second term, which "corrects" for $\rho^0$ meson virtuality,
is normalized to unity when
$\rho^0$ meson is on mass shell.
We expect the slope parameter of the order 
$B \sim$ 4 GeV$^{-2}$ and the parameter responsible for
off-shellness of $\rho^0$ mesons $\Lambda \sim$ 1 GeV.
We take the powers of the pomeron ($\alpha_{\Pom}$) 
and reggeon ($\alpha_{\Reg}$) terms from the 
Donnachie-Landshoff fit to the total $N N$ and $\pi N$
cross sections \cite{DL92}.
The parameters $C_{\Pom}$ and $C_{\Reg}$ are
obtained assuming Regge factorization and assuming that
$\sigma(\rho^0 \rho^0 \to \rho^0 \rho^0) =
 \sigma(\pi^0 \pi^0 \to \pi^0 \pi^0)$ 
(see e.g.\cite{SNS02}, they are:
$C_{\Pom}$ = 8.56 mb, $C_{\Reg}$ = 13.39 mb).
While this seems justified for the pomeron term,
it is not so obvious if it is true for the reggeon terms.
Consistent with our choice of normalization 
$\eta_{\Pom}({\hat s},{\hat t})$ and 
$\eta_{\Reg}({\hat s},{\hat t})$ are complex functions such that:
$\eta_{\Pom}({\hat s},{\hat t}=0) \approx i$
and $\eta_{\Reg}({\hat s},{\hat t}=0) \approx i + 1$.
Standard signature functions \cite{DDLN_book} are normalized
somewhat differently. The pomeron and reggeon trajectories
are parametrized as
$\alpha_{\Pom}(t) = 1.088 + 0.25t$ and
$\alpha_{\Reg}(t) = 0.5 + 0.9t$.

The differential cross section can be obtained from 
the corresponding amplitude as:
\begin{equation}
\frac{d \sigma_{\gamma \gamma \to \rho^0 \rho^0}}
{d \hat{t}} 
= \frac{1}{16 \pi \hat{s}^2} 
| {\cal M}_{\gamma \gamma \to \rho^0 \rho^0} |^2 \; .
\label{dsigma_dt}
\end{equation}
The total cross section $\hat{\sigma}$ can be obtained by 
integrating (\ref{dsigma_dt}) over $\hat{t}$
\begin{equation}
{\hat \sigma}_{\gamma \gamma \to \rho^0 \rho^0} =
\int_{t_{min}({\hat s})}
    ^{t_{max}({\hat s})}
\frac{d {\hat \sigma}}{d {\hat t}} \; d {\hat t} \; ,
\end{equation}
where $t_{min}$ and $t_{max}$ are ${\hat s}$-dependent
kinematical limitations of ${\hat t}$.

In Fig.\ref{fig:fit_gamgam_rhorho} we present
the corresponding $t$-integrated cross section together
with existing experimental data taken from 
\cite{MPW,original_data}.
Here the solid line includes the off-shell form factor
while the dashed line does not.
The vanishing of the VDM-Regge cross section at 
$W_{\gamma \gamma} = 2 m_{\rho}$ is due to 
$t_{min}$, $t_{max}$ limitations.
It is obvious from Fig.\ref{fig:fit_gamgam_rhorho} that the VDM-Regge 
model cannot explain the huge close-to-threshold enhancement. 
Up to now the origin of this enhancement remains unclear.
The VDM--Regge model nicely describes the experimental data for 
$W_{\gamma \gamma} \gsim 2.5 \, \mathrm{GeV}$.

\section{Results}

\begin{figure}[!h]    
\includegraphics[width=0.4\textwidth]{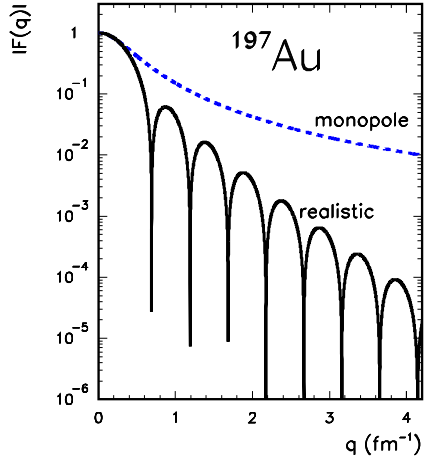}
   \caption{\label{fig:charge_ff}
   \small 
The modulus of the charge form factor $F_{em}(q)$ of 
the $^{197}Au$ nucleus for realistic charge distribution
(solid). For comparison we show the monopole form factor
often used in practical applications (dashed).
}
\end{figure}

The main ingredient for the calculation of photon
fluxes is the charge form factor of the nucleus.
In Fig.\ref{fig:charge_ff} we show the charge form factor
of $^{197}$Au calculated from
the realistic charge density as measured in 
electron scattering off nuclei \cite{VJV87}. 
One can observe many oscillations characteristic 
for relatively sharp edge of the nucleus. This form 
factor is used to calculate flux of equivalent photons 
according to Eqs.(\ref{flux}) and 
(\ref{E-field}). To illustrate the sensitivity
of our calculation on details of the form factor,
we also show a monopole form factor adjusted to the 
correct charge radius of the nucleus
$
F_{ch}(\vec{q}^2) = \Lambda^2 / (\Lambda^2 + \vec{q}^2) \, , 
$
with $\Lambda = 83 \, \mathrm{MeV}$.

\begin{figure}[!h]    
\includegraphics[width=0.4\textwidth]{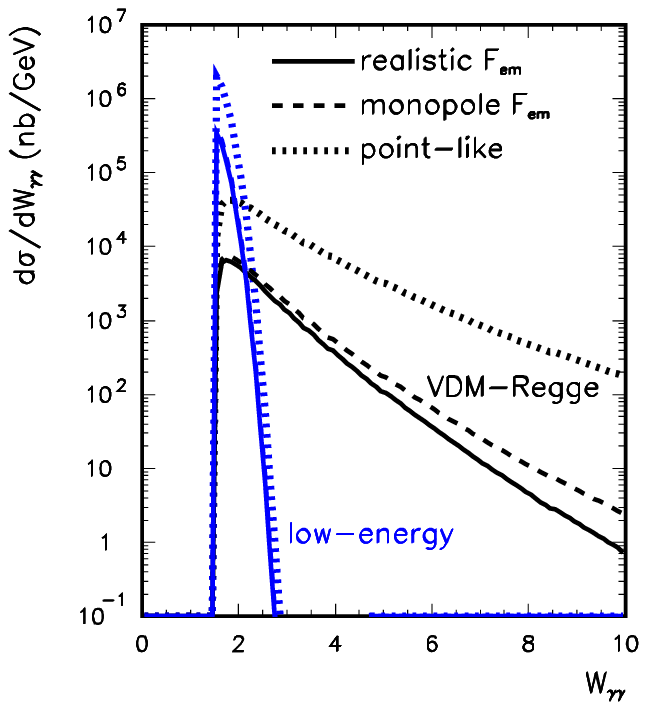}
   \caption{\label{fig:dsig_dW}
   \small 
The $Au + Au \to Au + Au + \rho^0 \rho^0$ cross section
as a function of $W_{\gamma \gamma} = M_{\rho \rho}$
for the RHIC energy $\sqrt{s}_{NN} = 200 \, \mathrm{GeV}$.
}
\end{figure}

In Fig.\ref{fig:dsig_dW} we show the distribution
of the cross section for the nucleus-nucleus scattering
in photon-photon center-of-mass energy $W_{\gamma \gamma}$ 
for both the low-energy component and high-energy 
VDM-Regge component. Below $W_{\gamma \gamma} = 2 \, \mathrm{GeV}$ 
the low-energy component dominates. The situation reverses above
$W_{\gamma \gamma} = 2 \, \mathrm{GeV}$. To study the 
high energy component one must impose an  extra cut on $W_{\gamma \gamma}$,
that is, the $\rho \rho$ invariant mass $M_{\rho \rho}$.
However, at RHIC energies the high--energy tail of the photon
spectrum is small, and the nuclear cross section drops quickly 
with increasing invariant mass of two-$\rho$ mesons.
By the solid lines we show the results for the realistic nuclear 
formfactor, while the dashed lines refer to the monopole form factor.
The results obtained with the different form factors start 
to diverge at $W_{\gamma \gamma} \gsim 4 \, \mathrm{GeV}$. This
is due to the increasing longitudinal momentum transfer involved
for higher invariant mass of the produced system. 
The $\gamma$-factors at RHIC are not too large, so that the
form-factor suppression is felt at such moderate invariant masses.
The result for the point--like nucleus, 
which we show by the dotted line, overestimates the realistic cross 
section by more than an order of magnitude.

\begin{figure}[!h]    
\includegraphics[width=0.4\textwidth]{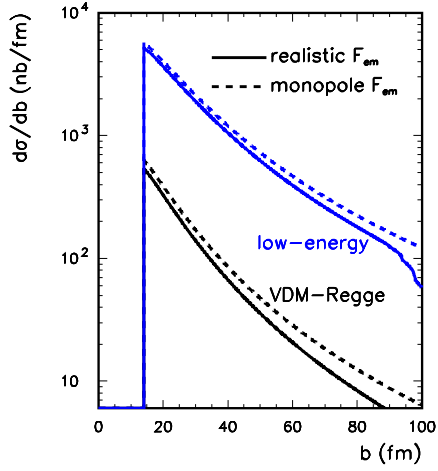}
   \caption{\label{fig:dsig_dbm}
   \small 
The $Au + Au \to Au + Au + \rho^0 \rho^0$ cross section
as a function of the impact parameter $b$ for 
$\sqrt{s}_{NN}$ = 200 GeV. 
}
\end{figure}

In Fig.\ref{fig:dsig_dbm} we show the  
distribution in impact parameter $b = |\bb_1 - \bb_2|$ 
(see also Fig.\ref{fig:collision_in_bspace}).
Again, we show distributions for the low- and high-energy
components separately. Also shown are the distributions
for the monopole form factor and for realistic charge 
density from \cite{VJV87}. 
One can clearly see different results for different 
approaches to calculate flux factors of equivalent
photons.
The sharp cutoff at $b = 2 R_A \sim 14 \, \mathrm{fm}$
is precisely the $\theta$--function coming from absorptive 
corrections.

\begin{figure}[!h]    
\includegraphics[width=0.4\textwidth]{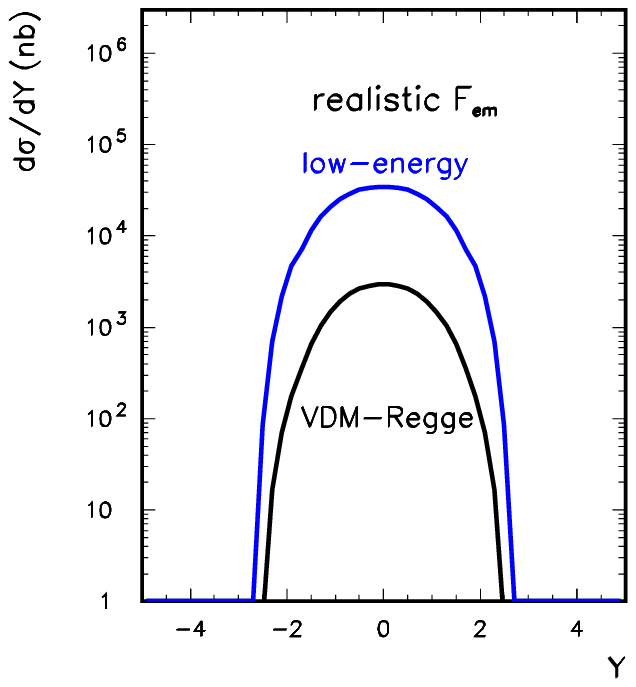}
\includegraphics[width=0.4\textwidth]{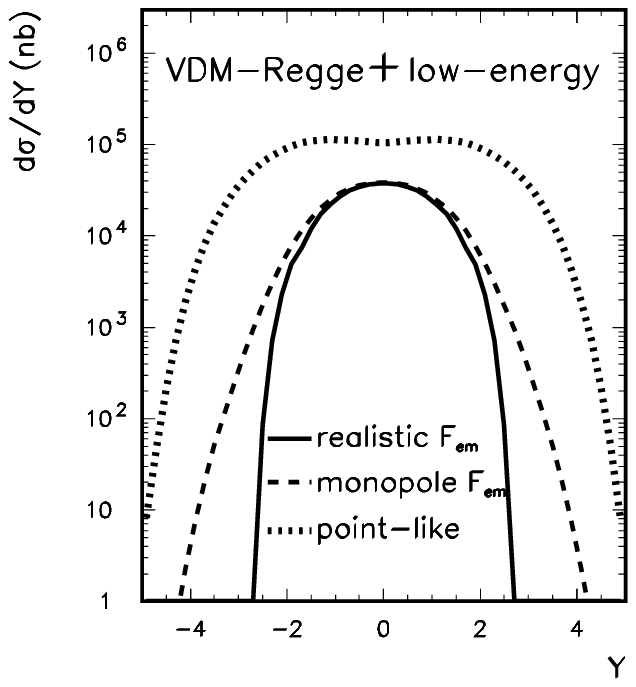}
   \caption{\label{fig:dsig_dY}
   \small 
The $Au + Au \to Au + Au + \rho^0 \rho^0$ cross section
as a function of the rapidity of the $\rho^0 \rho^0$ pair 
$Y$ for $\sqrt{s}_{NN}$ = 200 GeV.
In the left panel we show the decomposition into low-- and high--energy 
contributions. The right panel shows the sensitivity to the nuclear 
form factor.
}
\end{figure}

Finally in Fig.\ref{fig:dsig_dY} we show distribution
in rapidity--like variable $Y$. As far as small invariant
masses of $\rho^0 \rho^0$ pairs dominate, we may refer to it as
a rapidity of the $\rho^0 \rho^0$ pair.
Compared to the point-like case, the distribution 
obtained with realistic charge density is concentrated
at midrapidities, and configurations when both $\rho^0$'s
are in very forward or both $\rho^0$'s are in very backward
directions are strongly damped compared to the case
with point-like nucleus charges. 
At larger rapidities, one can again see 
a substantial difference between results obtained with 
an approximate monopole form factor  
and with the exact one calculated from realistic charge 
density. 

\section{Conclusions}

We have calculated, for the first time, realistic 
cross sections for exclusive $\rho^0 \rho^0$ production 
in ultrarelativistic heavy-ion collisions at RHIC 
in the framework of equivalent photon approximation.
We have discussed uncertainties related to the way
how the nuclear photon flux is calculated.
We have used realistic charge densities to calculate
the nuclear charge form factors.
The absorption effects have been included.
  
The low-energy part of the elementary 
$\gamma \gamma \to \rho^0 \rho^0$ process has been 
parametrized and the parameters have been fitted 
to the $e^+ e^-$ data while the high-energy part
has been modeled in the vector-dominance Regge type model
with parameters which are used to describe 
other hadronic processes. The model turned out to
be consistent with the highest-energy data points 
($W \sim$ 3-4 GeV) from $e^+ e^-$ collisions.

It was shown that a realistic calculation of both ingredients 
is necessary to make reliable estimates 
of the nucleus-nucleus exclusive production of the 
$\rho^0 \rho^0$ pairs.
Large cross sections, of the order of fraction of milibarn,
have been found. The bulk of the cross section is, however,
concentrated in low photon-photon energies 
(low $\rho^0 \rho^0$ invariant masses). Making cuts
on higher $\rho^0 \rho^0$ invariant masses one can 
easily select high-energy component.
The $\rho^0$ mesons, decaying into $\pi^+ \pi^-$, can 
be measured e.g. by the STAR detector at RHIC.
A Monte Carlo study is necessary to analyse feasibility
of a measurement of the process discussed here.

In the present analysis we have concentrated on processes
with final nuclei in the ground state. It is very 
difficult, if not impossible, to measure such very 
forward/backward nuclei. The multiple Coulomb 
excitations associated with $\rho^0 \rho^0$ production
may cause additional excitation of one or even both 
nuclei to the giant resonance region. The neutron
emission from the giant resonances can be used then 
to tag the processes.
The zero-degree calorimeters (ZDC) at RHIC can be used
to measure such neutrons.
We plan a detailed study of these processes in the future.

\vspace{1cm}

{\bf Acknowledgement}
We are indebted to W{\l}odek Guryn and 
Jacek Oko{\l}owicz for discussion.
This work was partially supported by the 
Polish Ministry of Science and Higher Education
under grant no. N N202 078735
and 1916/B/H03/2008/34.


\end{document}